\ifx\pdfoutput\undefined 

\else 
\documentclass[pdftex,12pt,a4paper]{article}
\usepackage[pdftex]{graphicx}
\DeclareGraphicsExtensions{.pdf,.png,.jpg,.mps}

\usepackage{hyperref}
\hypersetup{
  unicode=true,
  colorlinks=false,
  pdfborder={0 0 0}
}
\fi
\usepackage{color}
\usepackage{array}
\usepackage[utf8x]{inputenc}
\usepackage[T1]{fontenc}
\usepackage{enumerate}

\usepackage{amsfonts,amsthm,amssymb}
\usepackage{amsmath}
\usepackage{mathrsfs}

\begin{document}

\title{Stability of polytropic stars in Palatini gravity}
\author{Aneta Wojnar
\footnote{Center for Astrophysics and Cosmology \& PPGCosmo, Federal University of Esp\'irito Santo - Brazil}
\footnote{The author acknowledges financial support from FAPES (Brazil).}
}
\maketitle

\begin{abstract}
We will briefly discuss the necessary conditions for stability of polytropies in $f(\hat R)$ Palatini gravity and the differences with the General Relativity ones.
\end{abstract}

\section{Introduction}
Palatini $f(\hat R)$ gravity \cite{Palatini:1919di}, in the similar manner as 
other theories of gravity, usually is a background for astrophysical objects such as
black holes 
and neutron stars (NS). However, because of a degeneracy in the mass-radius profiles (the NS composition in the central core is still under debate), NS are not yet ideal objects to test theories of gravity. 

It turns out that non-relativistic stars, particularly dwarf ones, can also be used to test gravitational theories since the non-gravitational physics giving their
properties, such as metallicity and opacity for example, is not 
modified by them \cite{sak}. Thus, we would like to focus on the stability problem of such objects in the context of Palatini $f(\hat R)$ gravity.

\section{Palatini stellar objects}
\subsection{Relativistic and non-relativistic stars}
It was shown \cite{anet} that for a spherical-symmetric object,
whose matter is described by the perfect fluid energy-momentum tensor $T_{\mu\nu}$ and an equation of state given
by the barotropic relation $p=p(\rho)$, that the Tolman-Oppenheimer-Volkoff (TOV) equation can be written in the case of $f(\hat{R})$ Palatini gravity as
\begin{equation}\label{tov}
   \frac{d}{dr}\Big( \frac{\Pi(r)}{\phi^2(r)}\Big)=-\frac{\tilde{A}G\mathcal{M}}{r^2}\left(\frac{\Pi+Q}{\phi({r})^2}\right)\left(1+4\pi r^3\frac{\Pi}{\phi({r})^2\mathcal{M}}\right)
\end{equation}
where $\phi=f'(\hat{R})$, $\tilde A=\phi^{-1}(1-2G\mathcal{M}/r)^{-1}$ while $\Pi=\rho+\frac{c^4U}{2\kappa^2 G}$ and $Q=\rho-\frac{c^2U}{2\kappa^2 G}$ are generalized pressure 
and energy density, respectively, with $U=U(\phi)$ being a scalar field potential in the Einstein frame, depending on the $f(\hat{R})$ model.
It should be noticed that the coordinate $r$ is a conformal one thus the conformal transformation must be taken into 
account $r^2\rightarrow \phi r^2$.

Since Palatini gravity introduces modifications to the Newtonian hydrostatic equilibrium describing non-relativistic stellar 
objects such as dwarf stars or stars from the main sequence, applying non-relativistic limits to (\ref{tov}) $p<<\rho$ with
$4\pi r^3 p<< \mathcal{M}$ and $\frac{2G\mathcal{M}}{r}<<1$, one gets
\begin{equation}\label{part_EL}
 -r^2\Phi(r)p'=G\mathcal{M}(r)\rho(r),\;\;\;\;\mathcal{M}(r)= \int^r_0 4\pi \tilde{r}^2\frac{Q(\tilde{r})}{\Phi(\tilde{r})^2} d\tilde{r}.
\end{equation}

From now on, we will focus on the particular gravitational model, that is, the quadratic one $f(\hat{R})=\hat{R}+\beta \hat{R}^2$, while 
the matter part will be described by the polytropic equation of state, that is, $p=K\rho^{\Gamma}$.
In that case, it was shown \cite{fatibene} that the conformal transformation preserves the polytropic character of EoS in the Einstein frame. Thus, 
one simplifies further the mass function (\ref{part_EL}) to $\mathcal{M}(r)=\int^r_0 4\pi\rho \tilde{r}^2d\tilde{r}$. That makes possible to write down the generalized Lane-Emden equation 
for the quadratic model \cite{aneta3}
  \begin{equation}\label{leq}
\xi^2\theta^n\Phi^{3/2}+\displaystyle\frac{1}{1+\frac{\xi\Phi_{\xi}}{2\Phi}}\displaystyle\frac{d}{d\xi}\left(\displaystyle
\frac{\xi^2\Phi^{3/2}}{1+ \frac{\xi\Phi_{\xi}}{2\Phi}}\displaystyle\frac{d\theta}{d\xi} \right)=0,
\end{equation}
 where $\Phi=1+2\alpha\theta^n$, $\Phi_{\xi}=d\Phi/{d\xi}$, and
$
 \alpha=\kappa^2 c^2\beta\rho_c
$
with $\rho_c$ being the star's central density and $\kappa^2=8\pi Gc^{-4}$. The equation (\ref{leq}) possesses two exact solutions 
for $n=\{0,1\}$ \cite{artur}. The solutions (exact or numerical) allow to get 
the star's mass 
\begin{equation}\label{masa}
 \mathcal{M}=4\pi r_c^3\rho_c\omega_n,
\end{equation}
as well as central density, radius, and temperature, with $\gamma_n=(4\pi)^\frac{1}{n-3}(n+1)^\frac{n}{3-n}\omega_n^\frac{n-1}{3-n}\xi_R$, are
  \begin{equation}
 \rho_c=\delta_n\left(\frac{3\mathcal{M}}{4\pi R^3}\right),\;\;\;\;\;\
 R=\gamma_n\left(\frac{K}{G}\right)^\frac{n}{3-n}\mathcal{M}^\frac{n-1}{n-3}\xi_R,\;\;\;\;\;T=\frac{K\mu}{k_B}\rho_c^\frac{1}{n}\theta_n,
\end{equation}
where $k_B$ is Boltzmann's constant and $\mu$ the mean molecular weight.
However, $\omega_n$ and $\delta_n$ depend not only on the solutions of the LE eq. \cite{artur} but also on
 $\Phi$ and $\Phi_{\xi}$, such that
\begin{equation}
 \omega_n=-\frac{\xi^2\Phi^\frac{3}{2}}{1+\frac{1}{2}\xi\frac{\Phi_\xi}{\Phi}}\frac{d\theta}{d\xi}\mid_{\xi=\xi_R},
 \;\;\;\delta_n=-\frac{\xi_R}{3\frac{\Phi^{-\frac{1}{2}}}{1+\frac{1}{2}\xi\frac{\Phi_\xi}{\Phi}}\frac{d\theta}{d\xi}\mid_{\xi=\xi_R}}
\end{equation}

\subsection{Stability of polytropic stars in Palatini gravity}
The stability analysis for an arbitrary Lagrangian functional $f(\hat{R})$ performed in \cite{anet} shows that the stability condition of Palatini relativistic stars 
is similar to the one in GR: the condition depends on a given equation of state. Not very surprising, when Palatini gravity considered, one also needs to specify a 
model, that is, to examine the problem case by case.

In what follows, we would like to have a closer look on the above statement. Thus, let us limit ourselves to the non-relativistic hydrostatic
equilibrium equations to which we apply the polytropic equation of state such that we will use the Lane-Emden formalism to the stability analysis.

It is well-known (see e.g. \cite{cs}) that a necessary condition for stability can be expressed as
\begin{equation}
 \frac{\partial\mathcal{M}}{\partial\rho_c}>0.
\end{equation}
Applying it to (\ref{masa}) at the star's center with $\Phi=1+2\alpha\theta^n$ gives us an inequality of the form
\begin{equation}
 n-3-2\bar{\beta}(n+6)\rho_c-4\bar{\beta}^2(3+2n)\rho_c^2>0,
\end{equation}
while the equality will be satisfied by
\begin{equation}
 \rho_{c_1}=-\frac{1}{2\bar{\beta}},\;\;\;\;\rho_{c_2}=\frac{3-n}{3+2n}\rho_{c_1},
\end{equation}
where $\bar{\beta}=c^2\kappa^2\beta$. The first difference between GR and Palatini gravity is noticed immediately: the latter one introduces to the inequality a dependence on 
the model parameter 
$\beta$ apart from the polytropic parameter $n$ only as it happens in GR. The stability criterion crucially depends here on the sign 
of the parameter $\bar{\beta}$: For positive values of the parameter it may happen that stability occurs for the negative values of the central densities 
which is unphysical. Thus, in order to have $\rho_{c_2}$ positive and to be a stationary point, we immediately are left with the stable region for densities values from the range $(0;\rho_{c_2})$ 
but with ($\Gamma=1+1/n$)
\begin{equation}
 n>3\,\,\,\,\left(\Gamma<\frac{4}{3}\right).
\end{equation}
In GR we deal with unstable stellar configurations for such values of $n$; the Palatini model allows stable ones, though.

For the negative value of the parameter $\bar{\beta}$ the central density $\rho_{c_1}$ is a stationary point. In order for $\rho_{c_2}$ to be also a stationary point, the 
following condition must be satisfied for the polytropic index
\begin{equation}
 n<3,\,\,\,\,\left(\Gamma>\frac{4}{3}\right)
\end{equation}
giving also the range $(\rho_{c_2},\rho_{c_1})$ of the possible (and positive) central densities for which a star can be a stable configuration. However, in the case 
of the negative parameter $\bar{\beta}$, we will always
deal with some range of the stable configuration for each $n\geq0$, that is, $(0,\rho_{c_1})$, but with the only one stationary point $\rho_{c_1}$.

It should be mentioned that for the polytropic index $n=3$, which gives the EoS describing a white dwarf star, the stationary point exists only for the negative values of the 
parameter $\bar\beta$.

Let us notice that the stability analysis of the non-relativistic stellar objects allows to constraint the parameter $\bar{\beta}$: for example,
the typical density of a white dwarf is of the order $10^9$kg/m$^3$, therefore the parameter $\bar\beta\sim10^{-9}$m$^3$/kg.

\section{Conclusions}
The aim of this short letter was to demonstrate that in the case of Palatini quadratic model one deals with the similar situation as in GR, that is, the 
stability depends on the EoS. Though, as shown here for polytropic EoS, in the given gravitational model a necessary condition for stability allows 
to consider a wider range of polytropic index, being however dependent on the sign of the theory parameter $\beta$. A stationary point,
that is, a central density $\rho_c$ for which $\frac{\partial\mathcal{M}}{\partial\rho_c}=0$, for
polytropic white dwarfs ($n=3$) exists only for negative values of the parameter $\beta$. 

Furthermore, the stationary points obtained from stability analysis can also provide additional and independent of the theory of gravity, constraints for the model's
parameter because of the reasonable values of energy density which are given by the non-gravitational physics. Non-relativistic stars, although still having many secrets
to be revealed, are stellar objects much better known and understood than neutron stars, and thus giving interesting opportunities to test gravitational theories.

All together, this simple example shows that stability conditions should be reanalyzed in the context of modified gravity. 

\end{document}